\documentclass[a4paper,12pt]{article}

\usepackage{amsmath,amssymb,amsthm}
\usepackage[totalwidth=16.5cm,totalheight=23.5cm]{geometry}
\usepackage{graphicx}
\usepackage{comment}

\newcommand{\R}{{\mathbb R}}
\newcommand{\C}{{\mathbb C}}

\newcommand{\id}{{\mathbb I}}
\newcommand{\bR}{{\bar{R}}}
\newcommand{\ta}{{\tilde{\alpha}}}

\newcommand{\be}{\begin{eqnarray}}
\newcommand{\ee}{\end{eqnarray}}

\begin{document}
 \pagestyle{plain}
\title{Deformations of GR and BH thermodynamics}
\author{Kirill Krasnov \\ {}\\ {\small \it School of Mathematical Sciences, University of Nottingham, Nottingham, NG7 2RD} }
\maketitle
\begin{abstract}\noindent
In four space-time dimensions General Relativity can be non-trivially deformed. Deformed theories continue to describe two propagating degrees of freedom, as GR. We study Euclidean black hole thermodynamics in these deformations. We use the recently developed formulation that works with ${\rm SO}(3)$ connections as well as certain matrices $M$ of auxiliary fields. We show that the black hole entropy is given by one quarter of the horizon area as measured by the Lie algebra valued two-form $MF$, where $F$ is the connection curvature. This coincides with the horizon area as measured by the metric only for the case of General Relativity. 
\end{abstract}

\section{Introduction}

In four space-time dimensions, {\it complexified} General Relativity can be non-trivially deformed without changing its dynamical content -- the deformed theories continue to describe two propagating degrees of freedom. The first one-parameter family of deformations appeared in \cite{Capovilla:1989ac} in the context of "pure connection" formulation of gravity. This was later generalised to include an infinite number of deformation parameters -- the cosmological constants of \cite{Bengtsson:1990qg}. Bengtsson has studied these theories extensively in the early 90's referring to them as neighbours of General Relativity, see e.g. \cite{Bengtsson:1991bq}. The analysis was mainly carried out in the Hamiltonian framework. 

This infinite-parameter family of 4-dimensional gravity theories was rediscovered by the present author in \cite{Krasnov:2006du}. This work gave a Lagrangian Plebanski-type formulation with several types of auxiliary fields. It was later realised that a much more economical "pure connection" description is possible \cite{Krasnov:2011up}, \cite{Krasnov:2011pp}. This new development also emphasised the role played by the cosmological constant -- the pure connection description of both GR and its deformations works best for a non-zero value of $\Lambda$. 

The most recent development is work \cite{Herfray:2015fpa} that developed a "mixed" description in which independent fields are an ${\rm SO}(3)$ connection on space-time as well as a $3\times 3$ symmetric matrix of auxiliary fields. This formulation with auxiliary fields seems to be the easiest to study in practice, as it avoids unpleasant square roots appearing in the formulation  \cite{Krasnov:2011pp}. 

As we have mentioned, these deformations of GR are really theories of complexified gravity. Thus, the main dynamical object is an ${\rm SO}(3,\C)$ connection field satisfying certain second order PDE's. A certain complex metric is then constructed (algebraically) from the curvature 2-form of this connection. Various reality conditions can be imposed to select appropriate real slices. These depend on the desired signature. The easiest case is that of Riemannian signature metrics, when one simply works with ${\rm SO}(3,\R)$ connections. The difficult case is one of physical interest -- Lorentzian signature. Because on-shell the connection gets identified with the self-dual part of the Levi-Civita connection, in Lorentzian signature the connection must necessarily stay complex-valued. When the theory in question is General Relativity, it turns out to be possible to impose the reality condition stating that the metric (constructed from the connection) is real Lorentzian. This condition is compatible with (preserved by) the dynamics of the theory. This is no longer a good reality condition for the deformed theories. 

Thus, at present, it is not clear whether the deformed gravitational theories admit any physical interpretation. Still, they can be studied e.g. as interesting examples of modified Euclidean signature gravity. In fact, it is believed by most of the theoretical physics community that it is impossible to modify GR without changing its dynamical content (i.e. without adding new degrees of freedom) whatever the metric signature is. Such a statement is supported by various GR uniqueness theorems. The deformed gravity theories that are the subject of this paper, considered as theories of Riemannian signature metrics, give a concrete set of counterexamples to the common belief in GR uniqueness. Thus, these theories achieve the impossible, which is a sufficient motivation for their study. For more remarks on uniqueness and deformations of GR the reader is referred to \cite{Krasnov:2014eza}.

The purpose of this paper is to work out how the black hole thermodynamics gets modified by these deformed gravity theories. Even though black holes are intrinsically Lorentzian objects (surfaces that trap light are only possible if there are lightcones), one can and most often does study black hole thermodynamics by going to Euclidean signature. In this context, one can identify the on-shell value of the gravitational action with thermodynamic free energy, and its Legendre transform with the black hole entropy. It is this definition of the black hole entropy, which only refers to Euclidean signature metrics, that we extend to the deformed gravity theories in the present paper. 

It is well-known that the area-entropy relation is characteristic of Einstein's theory of gravity and gets modified in more general gravity theories. Without trying to give a fair review of this subject, we restrict ourselves to giving just one reference \cite{Banados:1993qp}, which considers Lovelock theories in higher space-time dimensions. The theories under study in the present paper are modified gravity theories, and so we not should expect that the entropy continues to be given by the horizon area. As we shall see, this expectation is correct in the sense that in general the entropy is no longer given by the horizon area as measured by the metric that signals the presence of the horizon. However, it turns out that the entropy is always given by some other simple geometric quantity that in a certain sense is a generalised notion of the horizon area. This quantity is sufficiently simple to be described already here. 

As we have already said, the main dynamical field in our description of GR and its deformations is an ${\rm SO}(3,\R)$ connection on a 4-dimensional manifold. We denote this connection 1-form by $A^i$, where $i=1,2,3$ is the Lie algebra index. Its curvature 2-form is given by $F^i = dA^i +(1/2) \epsilon^{ijk} A^j\wedge A^k$. The simplest to deal with, even though not most economical description, also involves a symmetric $3\times 3$ matrix  $M^{ij}$ of auxiliary fields. The action of the theory is then simply
\be\label{act-intr}
S[A,M] =\int M^{ij} F^i\wedge F^j,
\ee
and can be thought of as an integral of the Pontryagin density "twisted" by the matrix $M$. The matrix $M$ is not completely arbitrary, but is required to satisfy one condition that can be written as $g(M)=0$, with $g(M)$ being some ${\rm SO}(3)$-invariant function of $M$. It turns out that $g_{\rm EH}(M)={\rm Tr}(M^{-1}) - \Lambda$ corresponds to General Relativity, while changing $g(M)$ gives deformations of GR. 

A certain Riemannian signature metric can be constructed from the curvature 2-form $F^i$. This metric also depends on $M$. For the case of GR $g(M)=g_{\rm EH}(M)$, when the pair $A,M$ satisfy their field equations, the metric constructed can be shown to be Einstein with scalar curvature $4\Lambda$. For deformations of GR the metric constructed from on-shell pair $A,M$ is no longer Einstein. 

In this paper we consider the case $\Lambda <0$ so that the arising connections and metrics are asymptotically hyperbolic. Our restriction to non-zero $\Lambda$ is due to the fact that the description of gravity in terms of connections works best when $\Lambda\not =0$. Our choice $\Lambda<0$ is motivated by the availability in this case of an asymptotic region, where clean asymptotic boundary conditions can be imposed. It is possible to develop the $\Lambda>0$ story as well, but this will be complicated by the presence of another (cosmological) horizon. It is our desire to avoid these complications that motivates our choice of the sign. We note that a theory of asymptotically hyperbolic connections is recently developed in \cite{AH}. 

When the deformation away from GR is not too drastic, the obtained metric ends on a would-be conical singularity, as is the case in GR. Near the would-be conical singularity (Euclidean horizon) the solution looks like $S^2\times \R^2$. The condition of the absence of the conical singularity then gives the period of the Euclidean time (inverse temperature). We show that the black hole entropy is given by
\be\label{entr-intr}
{\cal S} = \frac{1}{4G} \int_{S^2} \sqrt{ |MF|^2}.
\ee 
Here $MF\equiv M^{ij}F^j$ is the restriction of the Lie algebra-valued 2-form $MF$ onto the sphere $S^2$ at the tip of the cone $S^2\times \R^2$. The norm under the square root is computed by taking an arbitrary volume form $dv$ on the $S^2$ and writing $M^{ij} F^j = u^i dv$. Then $\sqrt{ |MF|^2} := \sqrt{ u^i u^i} dv$ and does not depend on the choice of $dv$.  We note that a very similar type of object is used in loop quantum gravity in the construction of the area operator, see e.g. \cite{Ashtekar:1996eg}. Thus, our main result is that the black hole entropy in family of gravity theories (\ref{act-intr}) is $1/4G$ multiple of the area of the horizon as measured with respect to the 2-form $MF$. Only in the case of GR the area in (\ref{entr-intr}) coincides with the area as measured by the metric defined by $A,M$.

Interestingly, the 2-form $M^{ij}F^j$ is just the 2-form $B^i$ of Plebanski-like formulation \cite{Krasnov:2006du} of this class of theories. Thus, our result is that the entropy is just the quarter of the horizon area as is measured using the 2-form $B^i$. The 2-forms $B^i$ can be used to define yet another metric. Thus, one declares the 2-forms $B^i$ to be self-dual which gives the conformal class of the metric. This is the same conformal class as is defined by $F^i$, see more on this below. One can then define the conformal factor in such a way that the area of the spheres of symmetry is given by $\int \sqrt{|B|^2}$. With this definition of the metric, which is different from the one used in the main text to define the presence of the horizon, the entropy is quarter of the area. 

We present two calculations that lead to (\ref{entr-intr}). The first of these is a calculation of the thermodynamic free energy (on-shell gravitational) action for a certain one-parameter family of modifications. The Legendre transform of the free energy is then shown to give (\ref{entr-intr}). The other calculation is more general, and is applicable to any modified theory. Thus, we compute the boundary term that arises in varying the action (\ref{act-intr}) with respect to the connection. It is shown that this boundary term, when the boundary is taken to be the would-be conical singularity of the solution, coincides with (\ref{entr-intr}). Thus, the second more general derivation is in the spirit of \cite{Banados:1993qp}. It would be interesting to establish that the entropy (\ref{entr-intr}) is the one that appears in the first law of black hole mechanics. This is not attempted here and is left to future work. 

Our final comment is as follows. It is possible that the reader is not interested in any modifications of GR, especially considered here modifications with unclear or absent physical interpretation. For such a reader we mention that the analysis of this paper is still justified because we show how to solve $\Lambda <0$ Einstein equations in the language of connections, which is simple, but somewhat non-trivial exercise. We also obtain a new interpretation (\ref{entr-intr}) of the black hole entropy. 

We find it interesting that the entropy in deformed theories continues to remain so close to what it is in GR, essentially still being given by the quarter of the horizon area, with the properly defined notion of the area. This once again illustrates that deformations of GR are quite similar to GR in their properties. 

The organisation of this paper is as follows. First, in Section \ref{sec:theory} we give some more details on the gravity theories considered here, as well as on the description of GR in the language of connections. In Section \ref{sec:ss} we make the spherically symmetric connection ansatz and solve the arising field equations. The analysis in this section is completely general. Here we also obtain the boundary term arising in the variation of the action, to be identified with the entropy. We specialise to the case of GR in Section \ref{sec:GR} and to a one-parameter family of modified theories in Section \ref{sec:modif}. We compute the renormalised volume (thermodynamic free energy) and its Legendre transform (entropy) in Section \ref{sec:thermo}. 

\section{Einstein connections}
\label{sec:theory}

In this section we review how Einstein metrics can be described in terms of ${\rm SO}(3)\sim {\rm SU}(2)$ connections. We follow the recent description in \cite{Herfray:2015fpa}, where the reader is referred for more details. 

\subsection{The action functional}

Even though a "pure connection" description of non-zero scalar curvature Einstein metrics is possible, it appears that a description with a certain set of auxiliary fields is more useful. 

Let $A^i$ be an ${\rm SO}(3)$ connection on a 4-manifold, and $F^i = dA^i + (1/2) \epsilon^{ijk} A^j\wedge A^k$ be its curvature 2-forms. Let $M^{ij}$ be a symmetric $3\times 3$ matrix of auxiliary fields that is required to satisfy a certain condition, see below. Consider the following action
\be\label{action}
S[A,M] = \int {\rm Tr}( M F\wedge F).
\ee
The matrix $M$ is restricted to lie on a co-dimension one surface 
\be\label{surface}
g(M)=0
\ee
in the space of symmetric $3\times 3$ matrices, where $g(M)$ is a scalar-valued function. The surface that is relevant to Einstein connections/metrics is
\be\label{g-EH}
g_{\rm E}(M) = {\rm Tr}(M^{-1}) - \Lambda,
\ee
where $4\Lambda$ is the scalar curvature. The claim is that the critical points of (\ref{action}), with $M$ restricted to lie on (\ref{surface}) with (\ref{g-EH}) give rise to Einstein metrics.

\subsection{Euler-Lagrange equations}

Varying the action (\ref{action}) with respect to the connection we get
\be\label{eq-A}
d_A \left( M^{ij} F^j \right) =0.
\ee
Here we assumed that the manifold is compact, so that the integration by parts necessary to derive this field equation is justified. In the asymptotically hyperbolic case of relevance for us below, the bulk term in the action needs to be supplemented with a multiple of the Chern-Simons term on the boundary, see \cite{AH} and below. 

To get the equation for $M$, it is useful to impose the constraint (\ref{surface}) with the help of a Lagrange multiplier. To do this, we introduce an auxiliary 4-form field ${\mathcal V}$, to which we shall refer as the auxiliary volume form. We can then write
\be\label{X}
F^i\wedge F^j = X^{ij} \,{\mathcal V}.
\ee
The matrix $X^{ij}$ of curvature wedge products is defined, as ${\mathcal V}$ modulo multiplication by a function. We can then write the action as
\be
S[A,M,\mu] = \int_M \left( {\rm Tr}(M X) - \mu \, g(M) \right) {\mathcal V}.
\ee
The Euler-Lagrange equations for $M$ are then
\be\label{eq-M}
X = \mu \frac{\partial g}{\partial M},
\ee
which says that the matrix $X$ of curvature wedge products must lie in the direction orthogonal to the surface $g(M)=0$. 
When this equation can be solved, together with $g(M)=0$, for $M(X)$, the solution can be substituted to (\ref{eq-A}) to obtain the "pure connection" formulation of field equations, in which we obtain second-order PDE's for the connection. However, in many circumstances a more useful alternative appears to be to interpret (\ref{eq-A}) as a first-order PDE on the matrix $M$. Solving it one can then find $X$ from (\ref{eq-M}) and thus obtain a set of first-order PDE's on the connection components. It is this strategy that will be followed below in solving concrete problems in this setting. 

For the case of GR (\ref{g-EH}) the equation relating $X$ and $M$ takes the form $X=-\mu M^{-2}$. We can write it as $MX = -\mu M^{-1}$. Then, taking the trace and using the condition (\ref{g-EH}) we get
\be\label{eq-M-gr}
X = \frac{{\rm Tr}(MX)}{\Lambda} M^{-2},
\ee
which gives $X$ up to scale. 

\subsection{The metric}

We now describe the metric, constructed from the connection. This metric is Einstein when the connection satisfies the Euler-Lagrange equations (\ref{eq-A}), (\ref{eq-M-gr}).

First, one constructs the conformal metric. This is the unique conformal metric that makes the triple of curvature 2-forms $F^i$ self-dual. Explicitly, this is the conformal metric
\be
\sqrt{{\rm det(g)}} \, g_{\mu\nu} \sim \tilde{\epsilon}^{\alpha\beta\gamma\delta} \epsilon^{ijk} F^i_{\mu\alpha} F^j_{\nu\beta} F^k_{\gamma\delta}.
\ee

Second, one fixes the volume form ${\mathcal V}_m$ of the metric to be 
\be\label{volume}
- 2\Lambda \, {\mathcal V}_m = {\rm Tr}(M F\wedge F).
\ee
The numerical factor in this formula is adjusted to agree with some conventions introduced later. 

\subsection{Deformed Einstein connections}

An interesting class of deformations of the Einstein condition is obtained by simply changing the co-dimension one surface that restricts the matrix $M$ in the action (\ref{action}). This corresponds to changing the function $g(M)$ in the constraint equation $g(M)=0$. As we have reviewed above, the function that gives rise to Einstein connections is (\ref{g-EH}). Replacing $M$ by $M^{-1}$ in this function, i.e. $g_{inst}={\rm Tr}(M) - \Lambda$ turns out to give self-dual General Relativity, with solutions being half-flat Einstein metrics, see \cite{Herfray:2015fpa} for a discussion of this. More generally, deformations of Einstein connections can always be parametrised by making the cosmological constant in (\ref{g-EH}) a function of the trace-free part of the matrix $M^{-1}$:
\be
g(M) = {\rm Tr}(M^{-1}) - \Lambda(M^{-1}_{tf}), \qquad M^{-1}_{tf}\equiv\Psi:= M^{-1}- \frac{1}{3}  \id\, {\rm Tr}(M^{-1}),
\ee
so that
\be
M^{-1} = \Psi + \frac{\Lambda(\Psi)}{3} \id.
\ee
A particularly simple one-parameter family of modifications is obtained by taking
\be\label{alpha-family}
\Lambda(\Psi) = \Lambda_0 - \frac{\alpha}{2} {\rm Tr}(\Psi^2),
\ee
with $\alpha$ being a parameter of dimensions of inverse curvature. The modification of the Einstein condition that ensues becomes important in regions with large self-dual part of the Weyl curvature. 

In the deformed case, one continues to define the metric by the condition that the curvature 2-forms are self-dual. The volume form is then fixed to be a constant multiple of the form ${\rm Tr}(M F\wedge F)$, so that the action (\ref{action}) continues to receive the interpretation of the total volume of the space. 

\section{Spherically symmetric case}
\label{sec:ss}

\subsection{Equations}

We take the following spherically symmetric ansatz
\be\label{conn}
A^1 = a(R) dt + \cos(\theta) d\phi, \qquad A^2 = -b(R) \sin(\theta) d\phi, \qquad A^3 = b(R) d\theta.
\ee
The curvatures are
\be\nonumber
F^1= - a' dt\wedge dR + (b^2-1) \sin(\theta) d\theta\wedge d\phi, \\ \label{curv-ss}
F^2 = ab \, d\theta \wedge dt - b' \sin(\theta) dR\wedge d\phi, \\ \nonumber
F^3 = - ab \sin(\theta) dt\wedge d\phi + b' dR\wedge d\theta,
\ee
where the primes denote the derivative with respect to the radial coordinate, at this stage arbitrary.
The diagonal $X$ matrix can then be taken to be
\be\label{X-ab}
X^{11} = \frac{a'}{a}, \qquad X^{22} = X^{33}= \frac{b b'}{b^2-1}.
\ee
The field equations are obtained by substituting the above ansatz into the action (\ref{action}). We have
\be\label{eff-lagr}
S[a,b,M] = - 8\pi \beta \int dR \left( M_1 a' + 2 M_2 abb'\right),
\ee
where we integrated over the angular and time coordinates, assuming that $t$ is periodic with period $\beta$. By varying this action with respect to functions $a(R),b^2(R)-1$ we easily get the following system
\be
\left( M_1 a(b^2-1)\right)' = f(X) a(b^2-1), \qquad \left( M_2 a(b^2-1)\right)' = f(X) a(b^2-1),
\ee
where we manipulated the equations to map them into a form in which the right-hand-side is the same. Here
\be\label{fX-ss}
f(X)={\rm Tr}(M X) = M_1 \frac{a'}{a} + M_2 \frac{(b^2-1)'}{b^2-1}.
\ee
Of course the same equations can also be obtained from the equations $d_A M F=0$, but the above derivation from a simple action is more transparent. 

\subsection{Solution}

We now choose the radial coordinate so that
\be\label{radial}
f(X) a(b^2-1)=1.
\ee
The solution for the matrix $M$ is then immediately written down
\be
M_1 = f(X) (R+R_1), \qquad M_2 = f(X) (R+R_2),
\ee
where $R_{1,2}$ are integration constants. Alternatively, we can write
\be\label{M-soln}
M_1^{-1} = \frac{1}{f(X)(R+R_1)}, \qquad M_2^{-1}=\frac{1}{f(X)(R+R_2)}.
\ee

\subsection{Metric}

The metric is computed from the requirement that the curvature 2-forms (\ref{curv-ss}) are self-dual, and that the metric volume form is a constant multiple of the coordinate volume form. This last condition follows from our choice of the radial coordinate (\ref{radial}). We get the following metric
\be\label{metric-sp}
ds^2 = \sqrt{\frac{f(X)}{|\Lambda| X_1 X_2^2}} \left( X_1 \frac{a^2 b^2}{b^2-1} dt^2+ X_1 X_2^2 \frac{b^2-1}{b^2} dR^2 + X_2(b^2-1) d\Omega^2\right),
\ee
where as usual $d\Omega^2$ is the metric on the unit sphere. 

\subsection{Absence of the conical singularity}

Anticipating what will happen, let us assume that the function $b^2$ has a simple zero for some value $R=R_+$. Below we shall see that this is what happens in GR at the location of the Euclidean event horizon. This continues to be true in modified theories, and the purpose of this subsection is to establish the value of the angle deficit at the tip of the arising conical singularity. We can do this in full generality, without specifying the theory we work with. 

Thus, let us write
\be
b^2 = (b^2)'_+ (R-R_+) + \ldots.
\ee
Substituting this into (\ref{metric-sp}) we see that the correct radial coordinate near $R=R_+$ is given by
\be
\rho=\left( \frac{f(X) }{|\Lambda|X_1 X_2^2}\right)^{1/4} (X_1 X_2^2 (b^2-1))^{1/2} 2\sqrt{\frac{R-R_+}{(b^2)'_+}}.
\ee
Here all the terns are evaluated at $R=R_+$. We then substitute the $R-R_+$ found from this into the expression for the coefficient of the $dt^2$ term. We use the relation 
\be
\frac{1}{2}(b^2)' = (bb') = X_2(b^2-1)
\ee
to observe that all terms cancel leaving
\be
ds^2 = a^2_+ \rho^2 dt^2 + d\rho^2
\ee
as the metric near the tip of the cone. We learn that the correct period for the identification of the $t$ coordinate is
\be\label{beta-gen}
\beta = \frac{2\pi}{a_+}.
\ee
It is very interesting that the components of the connection seem to have direct physical meaning. Thus, at the horizon the component $b$ vanishes, while the component $a$ just measures the BH temperature!

\subsection{Boundary term and entropy}

The work \cite{Banados:1993qp} identified entropy with the boundary term of the Einstein-Hilbert action evaluated at the Euclidean horizon. Here we develop a similar interpretation for our theories. 

Thus, we first consider (\ref{eff-lagr}) and notice that in obtaining the associated Euler-Lagrange equations we have neglected the boundary terms 
\be\label{b-terms}
8\pi \beta \left( M_1 a + 2M_2 a b^2\right)
\ee
evaluated at both ends of the interval of integration over the radial coordinate. These terms at $R=R_\infty$ are taken care of by the Chern-Simons invariant that we will add at the boundary at infinity, see below. 

We now need to consider what happens at the point $R=R_+$. As we have seen, at this value of the radial coordinate $b^2=0$, and so the second term in (\ref{b-terms}) drops out. However, as we have also shown above, at this point we have (\ref{beta-gen}) and so the first term reduces just to
\be
16 \pi^2 M_1^+,
\ee
where $M_1^+\equiv M_1\Big|_+$.  If our expectation about the entropy being the boundary term of the action is correct, this quantity divided by $16\pi G$ must be the entropy 
\be\label{S-M1}
{\cal S}= \frac{\pi M_1^+}{G}.
\ee
This quantity can be written in a more invariant way as follows. Let us consider the 2-form $M^{ij} F^j$ restricted to the horizon 2-sphere. By inspection of (\ref{curv-ss}) we see that the only term that survives at the horizon is $F^1\Big|_+ = - \sin(\theta) d\theta\wedge d\phi$. This means that the entropy formula (\ref{entr-intr}) given in the Introduction is a covariant expression for our result (\ref{S-M1}). Below we will confirm that (\ref{S-M1}) is the correct expression for the entropy by explicitly computing the Legendre transform of the free energy for the one-parameter family of modified theories (\ref{alpha-family}).

\section{Solution in the GR case}
\label{sec:GR}

\subsection{Solving for the connection components}

We now specialise to the case of GR. We write the equation relating $M$ and $X$ as $MX=-\mu \, M^{-1}$ and take the trace of this equation. We get $-\mu = f(X)/\Lambda$, where we have used the definition (\ref{fX-ss}) and the condition ${\rm Tr}(M^{-1})=\Lambda$. But we can also take the trace of (\ref{M-soln}) to get
\be
\Lambda = \frac{1}{f(X)} \left( \frac{1}{R+R_1} + \frac{2}{R+R_2}\right) \equiv \frac{3R+2R_1+R_2}{f(X) (R+R_1)(R+R_2)}.
\ee
It is now convenient to shift the radial coordinate so as to impose
\be\label{R-origin}
2R_1+R_2=0.
\ee
We now take $R_1\equiv\bR$ to be the single parameter of the solution. This gives
\be
f(X) = \frac{3R}{\Lambda (R+\bR)(R-2\bR)}.
\ee
We now have $X=-\mu M^{-2}$, with $\mu$ expressed in terms of $f(X)$ that we just solved for. So, the final expressions for the quantities $X_1, X_2$ are
\be
X_1 = \frac{(R-2\bR)}{3R(R+\bR)}, \qquad X_2 = \frac{(R+\bR)}{3R(R-2\bR)}.
\ee
We now integrate the relations (\ref{X-ab}) to obtain the components of the connection
\be
a = K_1 \frac{R+\bR}{(3R)^{2/3}}, \qquad b^2-1 = K_2 \frac{R-2\bR}{(3R)^{1/3}},
\ee
where $K_{1,2}$ are integration constants. This gives a complete solution to the problem, modulo the issue of fixing (or interpreting) the integration constants $K_{1,2}$ and $\bR$.

\subsection{A relation between $K_1,K_2$}

There is a relation between the integration constants $K_{1,2}$ that follows from our gauge-fixing condition $f(X)a(b^2-1)=1$. Indeed, substituting the $f(X)$ that we have found above, as well as the connection components, we get
\be\label{K12}
K_1 K_2=\Lambda.
\ee

\subsection{Fixing $K_{1,2}$ from the asymptotics}

Let us fix $K_{1,2}$ from the large $R$ asymptotics of the metric (\ref{metric-sp}). We have $X_1,X_2=1/3R$ asymptotically, and $f(X)= 3/\Lambda R$. So, the conformal factor in (\ref{metric-sp}) behaves as
\be
\sqrt{\frac{f(X)}{|\Lambda| X_1 X_2^2}} = \frac{9R}{|\Lambda|} = 3l^2 R.
\ee
So, overall, for large $R$ the metric reads
\be
ds^2 = l^2\left( K_1^2 \frac{R^{2/3}}{3^{4/3}} dt^2 + \frac{dR^2}{(3R)^2} + K_2 \frac{R^{2/3}}{3^{1/3}} d\Omega^2\right).
\ee
We would like this to be the usual (asymptotically) hyperbolic metric
\be
ds^2 = l^2 \left( r^2 dt^2 + \frac{dr^2}{r^2} + r^2 d\Omega^2\right).
\ee
This means that asymptotically at least
\be
K_1^2 \frac{R^{2/3}}{3^{4/3}} = r^2, \qquad K_2 \frac{R^{2/3}}{3^{1/3}}  = r^2.
\ee
combining this relation with (\ref{K12}) we get
\be\label{R}
R=r^3 l^2,
\ee
and
\be
K_1 = \left(\frac{3}{l}\right)^{2/3}, \qquad K_2 = \left(\frac{3}{l^4}\right)^{1/3}.
\ee
Thus, finally, with these integration constants fixed from the asymptotics the connection components are
\be\label{ab-GR}
a =  \frac{R+\bR}{(Rl)^{2/3}}, \qquad b^2-1 = \frac{R-2\bR}{l(Rl)^{1/3}}.
\ee

\subsection{The final metric}

We can now write down the final expression for the metric. For the conformal factor we get
\be
\sqrt{\frac{f(X)}{|\Lambda| X_1 X_2^2}} = \frac{3l^2 R^2}{R+\bR}.
\ee
After numerous cancellations we get the following metric
\be
ds^2 = l^2 \left( b^2 dt^2 + \frac{dr^2}{b^2} + r^2 d\Omega^2 \right),
\ee
where (\ref{R}) is the relation that is valid everywhere. In terms of the radial coordinate $r$ the function $b^2$ takes the form
\be
b^2 = 1+ \frac{r^3 l^2-2\bR}{rl^2} = 1- \frac{2\bR}{r l^2} + r^2.
\ee
This becomes the usual Euclidean Schwarzschild AdS metric if we further replace $r\to r/l$ and identify
\be\label{MR}
\bR = M l,
\ee
where $M$ is the black hole mass.

\subsection{No conical singularity condition}

At the value $r_+$ of the radial coordinate satisfying
\be\label{rplus-GR}
1- \frac{2M}{l r_+} + r_+^2 = 0,
\ee
which is where the event horizon would be located in the Lorentzian signature solution, the metric has a conical singularity unless we identify time with the correct period. 

We already worked out the general formula for the inverse temperature $\beta$ in (\ref{beta-gen}), so we just need to apply the formula and see how the familiar AdS-Schwazschild solution temperature arises. The formula (\ref{beta-gen}) tell us that we just need to evaluate the value of connection component $a$ at the location of the event horizon. In terms of the $r$ coordinate the function $a$ (\ref{ab-GR}) is given by
\be
a = \frac{1}{r}\left(r^2 + \frac{M}{r l}\right).
\ee
Using (\ref{rplus-GR}) to express everything in terms of $r_+$ and substituting into (\ref{beta-gen}) we immediately get the familiar expression 
\be
\beta =  \frac{4\pi r_+}{1+3r_+^2}.
\ee
It is possible to parametrise the solution by $r_+$ rather than $M/l$ because, as it is easy to check, any positive value of $r_+$ can arise as $M/l$ varies between zero and infinity. Indeed, for $M\ll l$ $r_+=2M/l$, while for $M\gg l$ $r_+=(2M/l)^{1/3}$.

\section{A modified solution}
\label{sec:modif}

\subsection{The modified relation between $M$ and $X$}

We now solve all equations for the family of modified gravity theories (\ref{alpha-family}). The solution is essentially the same as in \cite{Herfray:2015fpa}, and so we simply list the relevant formulas. We define the matrix $R={\rm diag}(R+R_1,R+R_2,R+R_3)$. Then the matrix $X$ is given by formula (79) of \cite{Herfray:2015fpa}
\be
X = \frac{R^{-2}\left(\id - \frac{\partial \Lambda(\Psi)}{\partial \Psi}\right)}{{\rm Tr}\left[ R^{-1} \left(\id - \frac{\partial \Lambda(\Psi)}{\partial \Psi}\right)\right]}.
\ee
The matrix tracefree matrix $\Psi$ is given by 
\be
\Psi = \Lambda(\Psi)\left[ \frac{R^{-1}}{{\rm Tr}(R^{-1})} - \frac{1}{3}\id\right].
\ee
The only difference with the case considered in \cite{Herfray:2015fpa} is that now the second and third eigenvalues of all matrices are equal. 

\subsection{The solution for $\Lambda$}

As in \cite{Herfray:2015fpa}, we find the function $\Lambda(R)\equiv \Lambda(\Psi(R))$ by solving a certain quadratic equation. We introduce the function
\be
s_2(R) = \frac{1}{(R+R_1)^2} + \frac{2}{(R+R_2)^2} = \frac{3R^2+2R(R_2+2R_1)+R_2^2+2R_1^2}{(R+R_1)^2(R+R_2)^2}.
\ee
We make the same choice of the origin of the $R$ coordinate as in the GR case, namely (\ref{R-origin}), and set $R_1\equiv \bR$. We then have
\be
s_2(R) = 3\frac{R^2 + 2\bR^2}{(R+\bR)^2(R-2\bR)^2}.
\ee
We then introduce another function
\be
q(R) := \left(\frac{s_2}{s_1^2} -\frac{1}{3}\right) = \frac{2\bR^2}{3R^2},
\ee
which is non-negative and diverges at the singularity at $R=0$. The function $\Lambda(R)$ is then given as the solution of the quadratic equation
\be
-\frac{\alpha}{2} q(R) \Lambda^2 = \Lambda - \Lambda_0.
\ee
This gives
\be\label{lambda}
\Lambda = \frac{1}{\alpha q} \left( \sqrt{1+ 2\alpha \Lambda_0 q} - 1\right) = \frac{3R^2}{2\alpha M^2 l^2}\left( \sqrt{1- \frac{4\alpha M^2}{R^2}} - 1\right),
\ee
where we have substituted (\ref{MR}). As we shall see, this relation will still be valid because asymptotically we will have the solution behaving as Schwarzschild, with modifications becoming important only deep inside the manifold. We see that one must be careful here and for positive $\alpha$ make sure that $\alpha$ is not large enough so that the expression under the square root stays positive. 

\subsection{The solution for $X$}

Having determined $\Lambda(R)$, we can write down the solution for the matrix $X$ explicitly, this is formula (97) of \cite{Herfray:2015fpa}
\be
X = \frac{R^{-2}}{s_1(1+\alpha x \Lambda)}\left(1+\alpha\Lambda\left(\frac{R^{-1}}{s_1} - \frac{1}{3}\right)\right).
\ee
Spelling it out explicitly
\be
X_1 = \frac{R-2Ml}{3Ml} \left( \frac{1}{ \sqrt{R^2-4\alpha M^2}} - \frac{1}{ R+Ml} \right), \\ \nonumber
X_2 = - \frac{R+Ml}{6 Ml}\left( \frac{1}{\sqrt{R^2-4\alpha M^2}} - \frac{1}{R-2Ml}\right) .
\ee
We note that for $\alpha<0$ the singularity appearing in the case of GR at $R=0$ is absent in the modified theory. The components of the matrix $X$ stay finite at the would be singularity at $R=0$. There is a simple pole in $X_1$ located at $R=-Ml$, and a simple pole in $X_2$ located at $R=2Ml$, but these are present already in GR solution. These are coordinate singularities that do not correspond to any singularity in the connection components. 

\subsection{The solution for the connection components}

We can now integrate the relations (\ref{X-ab}) and obtain the connection components. Somewhat surprisingly, this can be done in a closed form. We get
\be\label{ab}
a = \frac{R+Ml}{(l/2)^{2/3}(R+\sqrt{R^2-4\alpha M^2})^{2/3}} \exp{ \frac{1}{3Ml} \left( \sqrt{R^2-4\alpha M^2} - R\right)}, 
\\ \nonumber
b^2-1=\frac{R-2Ml}{l(l/2)^{1/3}(R+\sqrt{R^2-4\alpha M^2})^{1/3}} \exp{ -\frac{1}{3Ml} \left( \sqrt{R^2-4\alpha M^2} - R\right)}.
\ee
We chose the integration constants so that we get the same asymptotics as the GR solutions (\ref{ab-GR}). Alternatively, the above solutions go to their GR counterparts as one takes the deformation parameter $\alpha\to 0$. For $\alpha<0$ there is no longer a singularity at $R=0$, with the connection components staying finite there. There is another region for negative $R$, which we do not analyse here as our main interest is thermodynamics. 

The fact that the Schwarzschild $r=0$ singularity gets resolved in modified theories of this type is not new, and has been studied (in a different formulation from the one considered here) in \cite{Krasnov:2007ky}. This reference also contains more information about the causal structure of the arising space times. We do not consider all these issues in the present paper, remaining in the Euclidean signature where the thermodynamic partition function can be computed. 

\subsection{The location of the tip of the cone}

The equation $b^2=0$ gives the following equation to solve for $R_+$ in terms of $M$
\be\label{R-M}
\exp{ \frac{1}{3Ml} \left( \sqrt{R^2_+-4\alpha M^2} - R_+\right)} = \frac{2Ml-R_+}{l(l/2)^{1/3}(R_+ +\sqrt{R^2_+-4\alpha M^2})^{1/3}}.
\ee
Unfortunately, it is no longer possible to solve for $M(R_+)$ explicitly.  But we can use this equation to get for $a_+$
\be
a_+ = \frac{2(Ml+R_+)(2Ml-R_+)}{ l^2(R_++\sqrt{R^2_+-4\alpha M^2})}.
\ee

\section{Renormalised volume and thermodynamics}
\label{sec:thermo}

\subsection{Renormalised volume}

The auxiliary matrix $M$ that appears in our Lagrangian is asymptotically a multiple of the identity matrix for any asymptotically hyperbolic solution, see (\ref{M-soln}). Because of this, the divergences appearing in the bulk part of the action (\ref{action}) are the same as those in a multiple of the Chern-Simons functional of the boundary connection. This means that the Chern-Simons invariant for the restriction of the connection to the boundary can be used to renormalise the bulk part of the action. 

We now compute the renormalised volume of the Euclidean BH metric, using the Chern-Simons invariant as the counter term necessary for the renormalisation. First, the bulk part is trivially computed thanks to our radial coordinate fixing condition. We have
\be\label{vol-sp}
\int {\rm Tr}(M F\wedge F) = - 2\int f(X) a(b^2-1) dR dt \sin(\theta) d\theta d\phi = -8\pi \beta (R_\infty - R_+),
\ee
where $R_\infty$ is the upper limit of the $R$ integration, and we have used the radial coordinate fixing condition (\ref{radial}). 

We now compute the CS functional for the connection (\ref{conn}). We write 
\be
dA^1 = da\wedge dt - \sin(\theta) d\theta\wedge d\phi,
\ee
and so then
\be\label{ada}
A^1 \wedge dA^1 = - adt\wedge \sin(\theta) d\theta\wedge d\phi + \cos(\theta) d\phi \wedge da\wedge dt.
\ee
The object $\cos(\theta)d\phi$ is not a globally defined 1-form, and so we cannot simply restrict $A^1 dA^1$ to the 3-boundary and integrate. Instead, let us rewrite the troublesome term as
\be
\cos(\theta) d\phi \wedge da\wedge dt = - d(adt\wedge \cos(\theta)d\phi) - adt\wedge \sin(\theta) d\theta\wedge d\phi.
\ee
We can now restrict to the 3-boundary and integrate. The integration gets rid of the total derivative term. Overall we get twice the contribution from the first term in (\ref{ada}), and not once as one might naively assume. Adding the contribution from the $(1/3) \epsilon^{ijk} A^i \wedge A^j \wedge A^k=2 A^1 \wedge A^2 \wedge A^3$ we get
\be
S^{CS} = 2 a(b^2-1) \int dt\wedge \sin(\theta) d\theta\wedge d\phi = 8\pi\beta a(b^2-1).
\ee
We can evaluate the quantity $a(b^2-1)$ e.g. from (\ref{ab}). We have
\be
a(b^2-1) = \frac{2(R+Ml)(R-2Ml)}{l^2(R+\sqrt{R^2-4\alpha M^2})}.
\ee
We can now expand this evaluated at $R=R_\infty$. We get
\be
-\frac{3}{\Lambda} a(b^2-1)\big|_{R_\infty} = R_\infty- Ml + O(1/R_\infty),
\ee
so that
\be\label{rv-sp}
-2 \Lambda RV=\lim_{R_\infty\to\infty}\left( \int {\rm Tr}(M^{-1}F\wedge F) + \frac{3}{\Lambda} S^{CS}_{R_\infty}\right) = 8\pi \beta(Ml-R_+).
\ee
Note that the CS functional does not simply remove the divergent term $R_\infty$ in (\ref{vol-sp}), but also adds a finite piece to the answer. This is precisely the same finite piece that appears in the literature, see e.g. \cite{Witten:1998zw}, formula (2.18). This finite piece is, however, usually obtained by considering the trace of the extrinsic curvature of the boundary. Here our Chern-Simons prescription gives it automatically. 

\subsection{Partition function}

The Euclidean partition function is the Euclidean Einstein-Hilbert action, which is a multiple of the volume. The Euclidean action is
\be
I = -\frac{1}{16\pi G} \int\sqrt{g}\, (R-2\Lambda) = - \frac{2\Lambda}{16\pi G} V,
\ee
where $V$ is the total volume. In the case of asymptotically hyperbolic manifolds the renormalised volume must be used. Substituting from (\ref{rv-sp}) we get
\be
I = \frac{\beta(Ml-R_+)}{2G}.
\ee

\subsection{A parametrisation}

It is not possible to solve the equation (\ref{R-M}) for either $R_+$ or $M$. However, we can rewrite it as a relation between 
\be
x=\frac{R_+}{M l}
\ee 
and $M/l$. We get
\be\label{M-x}
M/l =\frac{(x+\sqrt{x^2-4\ta})^{1/2}}{\sqrt{2}(2-x)^{3/2}} e^{(\sqrt{x^2-4\ta}-x)/2},
\ee
where
\be
\ta=\alpha/l^2.
\ee
For the case of GR the quantity $x$ changes $x\in(0,2)$, with small values corresponding to small black holes $M\ll l$ and values close to $2$ corresponding to large black holes $M\gg l$. One is not allowed to consider negative $x$ in the case of GR because of the square root in the numerator in (\ref{M-x}). However, the modified theory with $\tilde{\alpha}<0$ allows for negative $x$ as well, with large negative $x$ giving exponentially large masses. 

Using this parametrisation we can rewrite all the quantities as functions of $x$. We get for the inverse temperature
\be
\beta = \frac{\pi (x+\sqrt{x^2-4\ta})}{(M/l) (1+x)(2-x)} = \frac{\pi \sqrt{2}(2-x)^{1/2}(x+\sqrt{x^2-4\ta})^{1/2}}{1+x}e^{-(\sqrt{x^2-4\ta}-x)/2},
\ee
and for the partition function
\be
I = \left(\frac{l^2}{2G}\right) \beta (M/l) (1-x) = \left(\frac{l^2}{2G}\right) \frac{\pi(1-x) (x+\sqrt{x^2-4\ta})}{(1+x)(2-x)}.
\ee

\subsection{Thermodynamics}

It is now straightforward (but tedious) to check that
\be
\frac{\partial I}{\partial \beta} = \frac{\partial I/\partial x}{\partial \beta/\partial x} = \frac{Ml}{G}.
\ee
Thus, the mass parameter of our solution is indeed the thermodynamic energy (multiple of).

We can also compute the entropy
\be\label{entropy}
{\mathcal S} = \frac{\partial I}{\partial \beta} \beta - I = \left(\frac{l^2}{2G}\right) \frac{\pi(x+\sqrt{x^2-4\ta})}{2-x}.
\ee
This result for the entropy is rather simple. It is definitely not equal to the area of the sphere at $b^2=0$, as can be easily checked from the metric (\ref{metric-sp}).

\subsection{Interpretation}

The entropy obtained is no longer the horizon area, as it is in GR, but this is as expected for it is known that the entropy-area relation is characteristic of General Relativity. However, there is another way to characterise the gravitational entropy. Thus, as is shown in \cite{Banados:1993qp}, the entropy arises from the boundary term needed to make the Einstein-Hilbert action variational principle well-defined. This boundary term, computed around the would-be conical defect point of the Euclidean solution, is a multiple of the entropy. 

We have already computed the boundary term in (\ref{S-M1}). We will now show that (\ref{entropy}) coincides with (\ref{S-M1}). Indeed, from (\ref{M-soln}) we have $M_1=f(X) (R+\bar{R})$ with $f(X) = s_1(R)/\Lambda(\Psi)$. Thus, overall we have
\be
M_1 = \frac{3R}{\Lambda(R-2\bar{R})},
\ee
with $\Lambda$ given by (\ref{lambda}). Re-writing everything in terms of the quantity $x=R_+/Ml$ we have
\be
M_1^+ = \frac{l^2(x+\sqrt{x^2-4\tilde{\alpha}})}{2(2-x)}.
\ee
Comparing with (\ref{entropy}) we see that (\ref{S-M1}) indeed holds. 

We have obtained the expression (\ref{entropy}), interpreted as (\ref{S-M1}) for a particular one-parameter family of modifications of GR. However, the reasoning that led to (\ref{S-M1}) is general, and relies just on the fact that $b^2=0$ at the horizon. Thus, it can be expected that the entropy formula (\ref{S-M1}) and its covariant version (\ref{entr-intr}) hold for an arbitrary member of our modified family of gravity theories. 

\section*{Acknowledgments}

The author was supported by ERC Starting Grant 277570-DIGT\@. I am grateful to Joel Fine for many discussions on the subject of connection formulation of GR, in particular for a discussion about the spherically symmetric solution.

\end{document}